\documentclass[letterpaper]{article} 
\usepackage{aaai}  
\usepackage{times}  
\usepackage{helvet}  
\usepackage{courier}  
\usepackage[hyphens]{url}  
\usepackage{graphicx} 
\usepackage{natbib}  
\usepackage{caption} 
\usepackage[nolist]{acronym}

\acrodef{eeg}[EEG]{electroencephalography}
\acrodef{fnirs}[fNIRS]{functional near-infrared spectroscopy}
\acrodef{eda}[EDA]{electrodermal activity}
\acrodef{hmi}[HMI]{human-machine interface}
\acrodef{bci}[BCI]{brain-computer interface}
\acrodef{ai}[AI]{artificial intelligence}
\acrodef{emr}[EMR]{embodied musicking robot}

\title{\textit{Jess+}: designing embodied AI for interactive music-making}

\author {
    Craig Vear\textsuperscript{\rm 1},
    Johann Benerradi\textsuperscript{\rm 2}
}
\affiliations {
    \textsuperscript{\rm 1}University of Nottingham\\
    \textsuperscript{\rm 2}University of Cambridge\\
    craig.vear@nottingham.ac.uk, jb2615@cam.ac.uk
}

\begin{document}

\maketitle


\begin{abstract}
\begin{quote}
In this paper, we discuss the conceptualisation and design of embodied AI within an inclusive music-making project. The central case study is \textit{Jess+} an intelligent digital score system for shared creativity with a mixed ensemble of non-disabled and disabled musicians. The overarching aim is that the digital score enables disabled musicians to thrive in a live music conversation with other musicians regardless of the potential barriers of disability and music-making. After defining what we mean by embodied AI and how this approach supports the aims of the \textit{Jess+} project, we outline the main design features of the system. This includes several novel approaches such as its modular design, an AI Factory based on an embodied musicking dataset, and an embedded belief system. Our findings showed that the implemented design decisions and embodied-AI approach led to rich experiences for the musicians which in turn transformed their practice as an inclusive ensemble.
\end{quote}
\end{abstract}


\section{Introduction}
When T.S. Eliot wrote the words "Music heard so deeply that it is not heard at all, but you are the music while the music lasts" (T.S.Eliot's Four Quartets The Dry Savages), he was poetically describing the process a musician goes through as they become embodied. This process of embodiment happens when they 'speak through their instrument', and also when they become incorporated into the creative flow of the music with other musicians. This can happen to non-musicians too as they listen to music 'so deeply' that they become transported into the soundworld. This phenomenon of becoming embodied by, or through, music is at the centre of the research in this paper.

The central case study in this paper is \textit{Jess+} an intelligent digital score system for shared creativity with a mixed ensemble of disabled and non-disabled musicians. Through creating this digital score system we designed and implemented novel approaches to embodied AI and human-robot interaction. These design decisions are the central focus of this paper. We also discuss here the implementation of these designs into the case study and reflect upon the musicians' responses. As such, this paper comes as a complement to our related work focused on the user experience of \textit{Jess+} \cite{vear2024jess}, to give here the technical details necessary for the reproducibility of the system.

\subsection{Motivation}
The motivation to conduct this study came from a knowledge exchange discussion with an orchestra about how a digital score built using AI or bots can be used in real-time interactive music improvisation. [For the purposes of this paper a digital score is defined as a 'communications interface of musical ideas between musicians utilising the creative potential of digital technology' \cite{vear2019digital}]. This discussion was informed by a larger
European Research Council
funded research study into the transformation of the music score through digital technologies and its impact on musicianship and creativity (The Digital Score).
The frontier nature of this parent project involves seeking the boundaries of understanding and acceptance of the nature of a music score as a communications platform of transferable music ideas. Included in this is investigating the role of intelligent agents, and the interactive operationality of real-time intelligent scores. An important focus is how a digital score employing computational technology can be used as an inclusive vehicle for music-making, especially for musicians who face barriers to music-making and education. 

What emerged from the original discussion was a collaboration between Orchestras Live (a national producer creating inspiring orchestral experiences for communities across England) and Sinfonia Viva (an orchestra based in Nottingham, England) with the authors of this paper. It was designed as a knowledge exchange proof-of-concept in order to evaluate the impact and benefits of using AI and creative robotics to break certain barriers around disabled musician's access to creative music-making. The focus for the partners was using AI and the digital score concept to generate new modes of music-making and potential inclusive processes of creativity. The priority for the musicians was that it inspired creativity \textit{inside} music-making (an embodied activity) \cite{leman2007embodied}. The high-level research question for the researchers was: 

\textit{How to build an embodied-AI system that facilitates co-creation for an improvising ensemble of disabled and non-disabled musicians?}. 

\subsection{Basic description of the system}
\textit{Jess+} is an intelligent digital score system that uses AI and a robotic arm to amplify and communicate the creativity of an inclusive ensemble. The role of the arm is to present movements and gestures that inspire the musicians to make a sound and to co-create music through improvisation. The role of the AI is to sense the humans and to generate a response via the robot arm so that it is perceived as being meaningful in the flow of music-making \cite{bishop2018collaborative}, \cite{bailes2016musical}, \cite{bishop2019moving}. The sensing involves a microphone that listens to the sound produced by the human members of the ensemble and by direct \ac{eeg} and \ac{eda} input from the disabled musician. The overarching aim is that the digital score enables disabled musicians to thrive in a live music conversation with other musicians regardless of the persistent barriers of disability and music-making. 

\subsection{Summary of the findings}
Our findings showed that the design decisions that were implemented through the embodied AI approach led to rich experiences for the musicians which in turn transformed their practice and creative engagement as an inclusive ensemble.  They sensed that the AI and the robot were working with them inside the embodied world of music-making, and that its behaviour inspired actions and relationships that they perceived to be co-creative. However, many unknowns have emerged out of this proof-of-concept, and we can not say with any certainty \textit{how} they led to such transformations encounters for the musicians.


\section{Related work}

\begin{figure}
    \centering
    \includegraphics[width=0.6\linewidth]{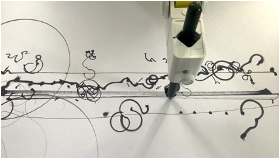}
    \caption{Example of notation}
    \label{fig:notation}
\end{figure}

There are several key concepts and research areas that this project addressed. In this section, we outline some of the areas that have conceptually informed this project and areas that we adopt and extend.

\subsection{Embodied AI \& musicking}
Understanding the nature of embodied AI gets us to the heart of the designs that are discussed here in this paper. Embodied AI can mean robotics (i.e. the AI controlling the robot, is expressed through the robotic hardware (a body)). But within the realm of real-time music-making, a body is only part of a communications mechanism between musicians \cite{bishop2019moving}. Another, arguably more important mechanism, is their interactions with others through their embodiment of the music-world \cite{nijs2009musical}.  It is within this realm that musicians build relationships, sense each other's musical play, and are incorporated together in the joint pursuit of meaning-making \cite{small1998musicking}.  So if we want AI/robots to join us inside the creative acts of music then how do we design and develop systems that prioritise the relationships that bind musicians inside the flow of music-making \cite{vear2022embodied}?

Two principles need explaining. First, that meaning in music is understood from the perspective of doing music. This is called \textit{musicking} and is defined by Christopher Small as "to music is to take part" \cite{small1998musicking}. Small wrote that taking part can happen "in any capacity, in a musical performance, whether by performing, by listening, by rehearsing or practising, by providing material for performance (what we call composing)".  Small stresses that "the act of musicking establishes in the place where it is happening a set of relationships, and it is in those relationships that the meaning of the act lies" \cite{small1998musicking}. 

Second, embodied AI has been defined as 'an intelligent agent whose operational behaviour is determined by percepts interacting to the dynamic situation within which it operates' \cite{vear2022embodied}.  This definition is built on three principles: 
    a) that artificial intelligence is not limited to the thinking-mind model;
    b) that the definition of AI adopts a modern approach as outlined by Russell \& Norvig \cite{russell2016artificial};
    c) that we understand meaning-making from the perspective of embodied cognition.
This clearly builds on Rodney Brooks's foundational work with behavioural AI and his coping machines, however, the dynamic environment that these 'creatures' are to exist within (and "do something in this world") is music \cite{vear2022embodied} \cite{brooks1991intelligence}.

\subsection{Musicking performance modelling}

The bulk of computational models for music performance (or "musicking", see above) attempt to codify musical expression in terms of mathematical formulae or symbolic programs. This type of modelling  \cite{cancino-chacon_computational_2018} concentrates on making machines perform music that can be perceived as being expressive. They accomplish this by translating artefacts of musicking, such as pitch, phrases, rhythms, and timing into expressive parameters (e.g. the "KTH rule system" for musical performance \cite{friberg_overview_2006}). These parameters are generally governed by weightings and a rule-based symbolic system of control which mostly does not work with the human musicians in-the-loop.

Most work in music performance modelling is aimed at classical, folk or jazz music \cite{cancino-chacon_computational_2018} \cite{lerch2021interdisciplinary} \cite{muller_automated_2010}. The most commonly modelled parameters within these systems are focused on loudness, articulation, ornamentation and timing, and generally using the MIDI protocol. MIDI is a useful platform as it enables the manipulation of musical parameters using the inbuilt control functions of velocity, pitch-bend, tuning. For simplicity, expressive parameters that are related are generally grouped together e.g., tempo and dynamics \cite{mcangus_todd_dynamics_1992}. Also, as these musics can be symbolically represented and shared through a music score, or some other forms of notation (such as ABC in folk music), lead sheets and common notational devices are a dominant visual platform for communications \cite{giraldo_machine_2016}. 

A common concern with music performance modelling is performance generation, rather than in-the-loop real-time interaction, for example, the \textit{MusicTransformer} system \cite{huang2018music}. This generates realistic accompaniments and performance given a melodic line input. Improvisation and variation are usually ignored when modelling classical music performance, but other styles (jazz and some folk traditions) consider them essential aspects of expressive performance.

\subsection{Musicking with robots}

Robotics is one of the ways technology is inviting itself into musical arts \cite{jeon2017robotic}. An excellent categorisation of approaches to musical robots is provided by Kemper \cite{kemper2021locating}, who describes 6 categories, that we here summarise into 3 main groups.

The first one consists of robot musicians of varying degrees of anthropomorphism, ranging from general-purpose robots playing music to robots specifically designed to play one instrument. They include for example the Toyota Partner Robot \cite{takagi2006toyota} a general-purpose humanoid robot able to play trumpet, Shimon \cite{hoffman2010shimon} the marimba player robot, or Haile \cite{weinberg2006robot} the robot percussionist.

The second main group involves the use of non-anthropomorphic robotic instruments that are not designed to resemble humans at all, but are instead designed as musical instruments themselves. Amongst them, we find for example MARIE \cite{rogers2015marie} an ensemble of monochord-aerophone robotics instruments or The Machine Orchestra \cite{kapur2011machine} a robotic orchestra of electromechanical instruments. On the most basic level, this also relates to the approach involving individual actuators used for their own sound production capabilities, which can for example be actuators from floppy discs and hard drives, that are used to produce sound directly.

Finally, the third group involves the use of cooperative musical robots that work together with humans to create music. These robots may be designed to play different instruments or parts of a composition and can communicate with each other to coordinate their actions. An example of this type of robot is Cyther \cite{barton2017cyther}, which allows a human and a robot to interact cooperatively through this robotic zither to generate music. This concept can even be extended to the augmentation of musicians through robotics, as it has been demonstrated for instance in the context of a robotic drumming prosthesis \cite{bretan2016robotic}.

\subsection{Wearables and human-machine interfaces (HMI) for music applications}

Wearable technologies have opened up new possibilities for music applications. They range from smartwatches to sensors embedded in garments, to neurotechnologies, and are a great medium to study how the body responds to music. For instance, skin conductance (also called galvanic skin response or \ac{eda}) has been linked to emotion induction through music \cite{khalfa2002event} \cite{ribeiro2019emotional}. This modality was for example used towards making a \ac{hmi} generating music based on the emotional response \cite{daly2015towards}, or to create music recommendation systems \cite{ayata2018emotion}. Heart rate variability as measured by smart watches has also been used to create personalised music recommendation systems thanks to decision tree \cite{chiu2017develop}. Other studies also demonstrated how biofeedback using this modality was used to reduce music performance anxiety \cite{thurber2010effects}.

Neurotechnologies play a key role in \ac{hmi}s \cite{singh2021developments}. Many non-invasive technologies have been used for example with \ac{fnirs} where support vector machine classification was used for a music learning \ac{bci} \cite{yuksel2016learn}. Of course \ac{eeg} as one of the most popular modalities for non-invasive \ac{bci}, has been used to develop brain-computer music interfaces for music generation \cite{miranda2008towards}. This type of interface has been used for music-making by people with severe disability \cite{miranda2011brain}. It is also shown that \ac{eeg} has also great potential in passive brain-computer interfaces (\ac{bci}s) for interactive music performances \cite{belo2021eeg}. \ac{bci}s can be divided into 2 categories \cite{clerc2016brain}: an active \ac{bci} is a system where the user is voluntarily controlling an effector, while a passive \ac{bci} relies on involuntary generated input that is used to control an effector. Such interfaces are typically built using machine learning to make predictions from brain signals \cite{clerc2016brain} \cite{naseer2015fnirs}, and therefore require to be trained on datasets.

Finally, applications of  \ac{hmi}s  are also very developed in robotics \cite{berg2020review}. As a result, we here pose the challenge to build a \ac{hmi} with multi-modal wearable and audio inputs that interacts with its environment through a robotic arm. To the best of our knowledge, such an embodied system has never been built so far. The following section \ref{design} is dedicated to the description of that system.


\section{Design of \textit{Jess+}}
\label{design}
In designing the system for \textit{Jess+}  it was important for the team to define the following objectives:
\begin{itemize}
    \item \textit{Obj 1}: Develop an agile methodology that placed user-experience at the centre of the developmental process.
    \item \textit{Obj 2}: Design a stack architecture that mirrored Brook's notion of subsumption that adopt his behavioural principles of a 'coping creature' with suitable translations for the embodied realm of musicking \cite{brooks1991intelligence}. Furthermore, that each module in the stack architecture could be replaced and updated, without reworking the whole system.
    \item \textit{Obj 3}: Develop an interactive loop that co-operated in real time within the dynamic world of musicking \cite{brooks1991intelligence}.
    \item \textit{Obj 4}: Develop a percept / sensor input layer that connected the disabled musician into the AI stack architecture and incorporated the non-disabled musicians into the system.
    \item \textit{Obj 5}: Design a musicking-specific approach to the implementation of machine learning that operated within the dynamic world of musicking.
    \item \textit{Obj 6}: Build a response system that could be trusted and was dynamic to the real-time changes in the music-world.
    \item \textit{Obj 7}: Develop a belief system that embedded an aesthetic design and musical focus developed by the 3 musicians, that was trustworthy and helped to create meaningful interplays with the human musicians.
\end{itemize}         

These objectives built on the existing research into embodied musicking robots \cite{vear2021creative} \cite{vear2022embodied} and served as a work plan with which to break the main challenge into manageable packages.

\subsection{User centred design}
The development of the software was open-source and incremental. We employed an Agile workflow for continuous delivery where new modifications to the system would be reviewed to ensure quality, in order to maintain a working version at all times. This was useful to always have the system ready for the musicians to test on multiple occasions (described in \ref{workshops}). Their feedback was critical and was taken into account through each testing session following a user-centred design for improving the system at each new iteration.

\subsection{Interaction loop: what is it doing?}
\begin{figure}
    \centering
    \includegraphics[width=\linewidth]{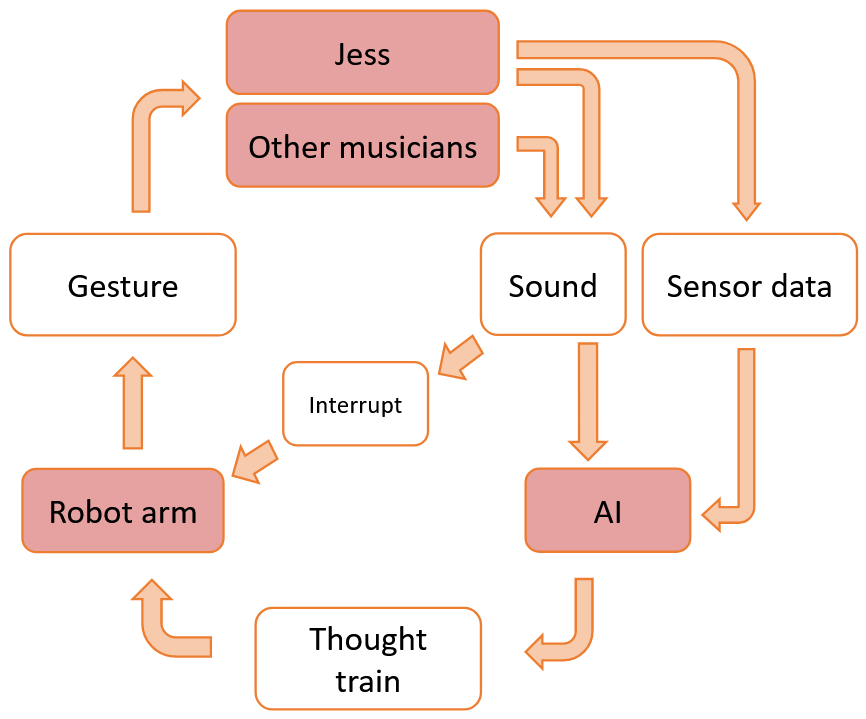}
    \caption{Interaction design \cite{vear2024jess}}
    \label{fig:interaction-design}
\end{figure}

\textit{Jess+} is based on a closed-loop interaction design (Figure \ref{fig:interaction-design}) that we cover in details here. Starting at the top right corner of Figure \ref{fig:interaction-design} [Sound \& Sensor data] sound from all the musicians (including Jess, the disabled musician) and sensor data from Jess, are processed by the AI Factory [AI] (discussed in \ref{ai-factory}), which produces a [Thought train] value, used to determine a robotic gesture selected from a pre-defined/ composed library of gestures (discussed in \ref{language}) that the robot arm will perform. Those gestures are then interpreted by all the musicians who play music improvising, and Jess reacts physiologically to it, closing the loop. That main loop can also be bypassed when the sound is too loud, interrupting the robot's gestures and moving on to a new main cycle (called the startled response). This creates an interaction between the musicians and the robot arm going both ways: on the one hand, the robot is "dancing" to what the musicians produce (sound and sensor data); on the other hand, the robot is acting as a "conductor" that the musicians can take inspiration from. This enables to blur the line between passive and active \ac{bci}, as musicians can voluntarily try to influence the robot, or let it go with the flow. In this system, the AI and the robot arm are not an assistive tool, but rather co-creative other.

\subsection{Modular design: how does it work?}
\textit{Jess+} was built on a modular system comprising high-level layers (also outlined in \cite{vear2024jess}):
        \begin{itemize}
            \item \textit{Layer 1} - percept input and formatting. This module manages and formats all the real-time input data and sends it to the AI factory and gesture manager for processing.
            \item \textit{Layer 2} - AI Factory. This module generates streams of data from the 7 neural network models that are housed in the AI factory. Its purpose is to generate a constant flow of predicted data from each model (emulating a busy mind negotiating options inside music-making).
            \item \textit{Layer 3} - gesture manager. This module chooses one of the outputs from the AI factory (layer 2), or live input stream (layer 1) and holds this stream for a few sections before randomly choosing the next (we called these "trains of thought"). There is a 'startled' function that interrupts these trains when the live sound reaches a certain threshold.
            \item \textit{Layer 4} - belief system, robot choices and language. This module takes the single output stream from the gesture manager (layer 3), and uses its value to determine which of the pre-defined movement gestures the arm is to conduct, and also decides operational factors like speed, acceleration etc.
        \end{itemize}

The system was divided into multiple threads working simultaneously and sharing data states using a hive mind data structure following a \textit{Borg} pattern \cite{martelli2005python}: a thread processing the audio input from the microphone; a thread processing the sensor data; a thread for running the AI Factory; a thread managing robot gestures; a thread saving audio and thought train data for generating an artefact.

\subsubsection{Sensors and equipment}
Multiple sensors enabled the system to be aware of its environment. First of all, a microphone (Røde NT-USB) streams the real-time ambient sound to capture music played by the musicians positioned around it. In addition, physiological data sensors are worn by the disabled musician, including \ac{eeg} (BrainBit with 4 sensors measuring brain activity in the temporal and occipital lobes) and \ac{eda} (using a BITalino placed on the hand palm).

The robots used in this study were the Magician Lite from Dobot or xArm 6 from UFACTORY. Cartesian positions of the robot arm tool centre were captured in 3 dimensions (x and y on the horizontal plane and z as the height). In the case of the Magician Lite, it was positioned on a table with a sheet of paper, and a pen is attached to it. In the case of the xArm 6, it was screwed on a pallet on top of a drawing board, and either had a feather attached to it, or an array of 4 pens of different colours that the gesture manager can choose from during the piece for drawing on a sheet of paper positioned on the board. For both robots, a fenced area (safety zone) was encoded, and the musicians took place outside of it.

\subsubsection{Language}
\label{language}
A gestural language was implemented in \textit{Layer 4} of the system. This was a repository of fixed movement types that outlined the aesthetic nature of \textit{Jess+} as a digital score (as opposed to an open drawing machine) and also enabled the musicians to contribute to this aesthetic by suggesting movements and inspiration points into this repository (see below). Immediate parameters such as point location, speed, acceleration and size, for example, were chosen at random when the AI had made a decision about which movement gesture to use at a given moment.

We chose to see this language as a \textit{belief system}, and as a way of embedding a focused understanding of what music is, rather than knowing everything about music and every gesture possible. This \textit{belief system} attuned the AI's responses to a specific aesthetic, and aligned to the core definition of a digital score (above), as a communications package of a musical idea. Other interactive response parameters such as behaviours, weightings,  speed range, and timings were considered to be part of this \textit{belief system}. This aligns with Barr et al.'s definition of robotic belief as "an acceptance that something is true, or that it has trust or confidence in something" \cite{barr1981handbook}.  

This language used in this digital score is a set of predefined symbolic gestures inspired by:
\begin{itemize}
    \item basic shapes such as squares, triangles, circles and starbursts;
    \item drawing/movement gestures inspired by Cardew's composition \textit{Treatise} (1967)  \footnote{\url{https://medium.com/nightingale/treatise-a-visual-symphony-of-information-design-2ced33ef01a0}};
    \item drawing/movement gestures inspired by Wolff's composition \textit{For 1, 2, or 3 people} (1964) \footnote{\url{https://spiralcage.wordpress.com/2011/03/06/music-for-merce-part-2/christian-wolff-for-12-or-3-people}};
    \item off-page 3-dimensional gestures across the fenced space of the robot arm's movement.
\end{itemize}

\subsubsection{AI Factory}
\label{ai-factory}
An AI Factory was trained in a way that attempted to encapsulate correlations between bodily movement, physiological response and music invention. The dataset used in the training is introduced below. An important aspect of the conceptualisation of the AI factory was that rather than training a single multi-variable neural network to predict this trinity of "mind", "body" and "music" in one calculation, it would produce multiple "trains of thought" using a factory of simple neural networks. This is a poetic model of the mind as proposed by Gelernter \cite{gelernter2010muse}, and formed another aspect of the AI's \textit{belief system}, but also aligned to personal experiences of the principal investigator through many decades of professional music practice.

\begin{figure}
    \centering
    \includegraphics[width=\linewidth]{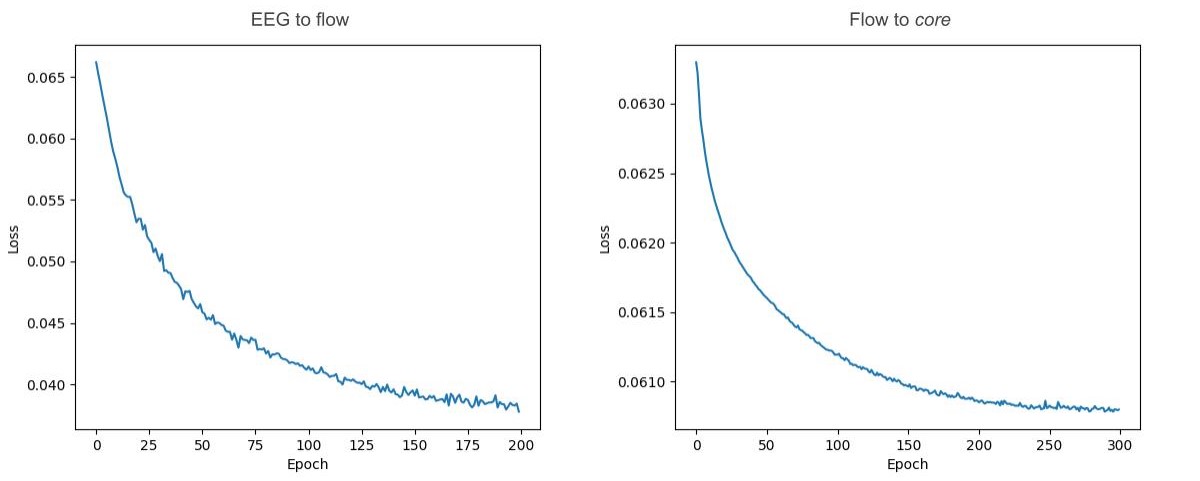}
    \caption{AI Factory validation loss (mean squared error) with respect to the epoch number: \textit{left side:} \ac{eeg} predicting flow (learning rate: 5e-5, batch size: 32); \textit{right side:} flow predicting \textit{core} (learning rate: 1e-5, batch size: 16)}
    \label{fig:training-graphs}
\end{figure}

\paragraph{Dataset}
A set of machine learning models that we call the AI Factory was trained on the Embodied Musicking Dataset \footnote{\url{https://rdmc.nottingham.ac.uk/handle/internal/10518}} capturing human creativity from inside the embodied relationships of music performance. This dataset was collected in 2020 in
New Haven (USA) and Leicester (UK), 
and consisted of recordings of pianists (N=8) improvising on a jazz backing track, each for 3 separate performances of 5 mins 34. This study was approved by the
De Montfort
The university ethics committee and all the participants gave informed consent before taking part.  Data was recorded during their performances including: video recording with audio, skeleton positions measured with an Intel RealSense camera, \ac{eda} measured with a BITalino and \ac{eeg} measured with a BrainBit. All the data was recorded at 10 Hz except the video (30 frames per sec) and audio (44,100 Hz). After each session, musicians were asked to watch the recording of their performance and rate their "flow" level by adjusting a cursor continuously on a software throughout the whole replay. This "flow" measure was their subjective evaluation of how deeply involved \cite{nijs2009musical} they were in the music (sometimes called groove) and was also recorded at 10 Hz.

\paragraph{Feature extraction}
The following features were then extracted:
\begin{itemize}
    \item the audio envelope using a Hilbert transform, downsampled to 10 Hz to match the rest of the data (1 channel);
    \item the x and y positions of the \textit{core} (2 channels), the midpoint between the two shoulder skeleton points;
    \item the \ac{eeg} data, filtered using an infinite impulse response Butterworth high-pass filter with a cut-off frequency of 1 Hz to remove slow drifts (4 channels);
    \item the raw \ac{eda} (1 channel) and flow values (1 channel), without further processing.
\end{itemize}
The resulting time-series data was then cut into 5-second chunks for a total of 50 time points for each example.

\paragraph{Models}
\label{models}
7 deep learning models were trained on these extracted features to predict features from each other (we reference them here with numbers) :
\begin{enumerate}
    \item[1)] Hilbert transform of the audio predicting "flow" assessment (by the musician);
    \item[2)] Hilbert transform of the audio predicting \textit{core} mid-shoulder position;
    \item[3)] \textit{core} mid-shoulder position predicting "flow" assessment;
    \item[4)] filtered \ac{eeg} predicting "flow" assessment;
    \item[5)] raw \ac{eda} predicting "flow" assessment;
    \item[6)] "flow" assessment predicting \textit{core} mid-shoulder position;
    \item[7)] "flow" assessment predicting Hilbert transform of the audio;
\end{enumerate}

All those models adopted a convolutional encoder-decoder hourglass architecture \cite{milletari2016v} (encoding a feature and decoding, another feature), composed of (in order, from input to output): a batch-normalised one-dimensional convolutional layer with \textit{n} input channels (with \textit{n} number of channels of the feature) and 8 output channels, a kernel size of 5 (one-dimension kernel) and a stride of 3, with a ReLu activation; a batch normalisation; a flattening step; a fully connected layer of 32 neurons with ReLu activation; a fully connected layer of 128 neurons with ReLu activation; an un-flattening step; a transposed convolutional layer with 8 input channels and \textit{n} output channels (with \textit{n} number of channels of the feature), a kernel size of 5 and a stride of 3, with a sigmoid activation function.

\paragraph{Training}
The dataset was randomly split into training (75\%) and validation (25\%) sets. Data was normalised with min-max feature scaling by computing the minimums and maximums for each channel of each feature across the training set. The training was then performed using a mean squared error loss with Adam optimiser \cite{schmidt2020descending}. 

For this proof of concept, hyperparameters (learning rate, batch size, and number of epochs) were roughly optimised manually with the help of training graphs, ensuring a mean squared error decrease on the validation set and training was stopped after the performance stopped improving noticeably (Figure \ref{fig:training-graphs}). The trained models were then saved in order to be deployed in \textit{Jess+}.

\paragraph{Deployment}
The trained models are loaded when running \textit{Jess+} and are used to compute predicted features, later on processed for generating robot arm gestures, using live inputs from the sensors.

For this purpose, 5-sec buffers of data from the sensors are stored using a sliding window, and those are processed similarly to the features from the dataset, generating the following:
\begin{itemize}
    \item the audio envelope is extracted from the microphone input buffer using a Hilbert transform, downsampled to 10 Hz to match the rest of the data (1 channel);
    \item the x and y positions of the robot arm tip (2 channels), robotic representation of the core mid-shoulder position;
    \item the \ac{eeg} data buffer, detrended to remove slow drifts (4 channels);
    \item the \ac{eda} data buffer, detrended to remove slow drifts (1 channel);
\end{itemize}

\begin{figure}
    \centering
    \includegraphics[width=\linewidth]{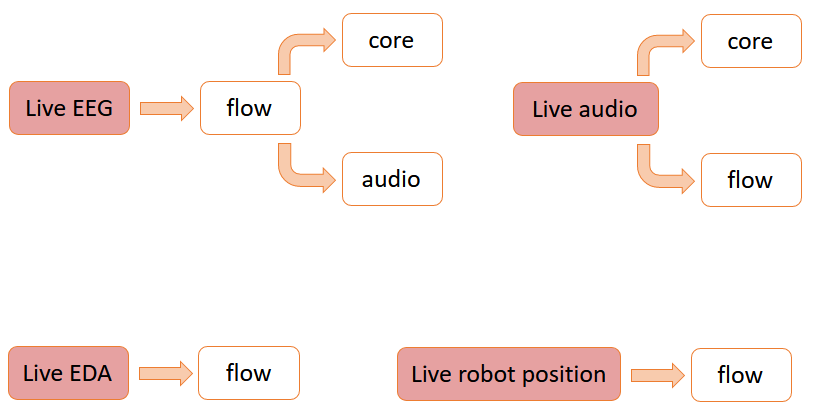}
    \caption{AI Factory \cite{vear2024jess}}
    \label{fig:ai-factory}
\end{figure}

All those features are normalised in real-time across the 5-sec buffer (as opposed to the whole training set) for more dynamic response and fed as input of the models trained on the Embodied Musicking Dataset. They are used in the following way (Figure \ref{fig:ai-factory}):
\begin{enumerate}
    \item[a)] live audio envelope is fed into model 1) to output a predicted "flow";
    \item[b)] live audio envelope is fed into model 2) to output a predicted \textit{core};
    \item[c)] live robot arm positions are fed into model 3) to output a predicted "flow", this is to emulate the concept of self-awareness;
    \item[d)] live \ac{eeg} is fed into model 4) to output a predicted "flow";
    \item[e)] live \ac{eda} is fed into model 5) to output a predicted "flow";
    \item[f)] predicted "flow" from d) is fed into model 6) to output a predicted \textit{core};
    \item[g)] predicted "flow" from d) if fed into model 7) to output a predicted audio envelope. 
\end{enumerate}

\subsubsection{Gesture manager}
From the AI Factory running in \textit{Jess+}, each model output is averaged across all dimensions to generate 7 thought train streams of values between 0 and 1, corresponding to the 7 models (above), at a 10 Hz frequency in real-time. To those are added to 2 other thought train streams: the normalised audio amplitude, and a stream of random values (\textit{random poetry}) between 0 and 1.

A gesture manager determined gesture phrases of 3 to 8 seconds (selected randomly) with the robot arm. These broad fuzzy parameters were part of the \textit{belief system} and were fine-tuned to work effectively with the musicians. Its function was to select one of the 9 streams from the AI factory to be used for robot arm movement selection: the audio stream is listened to with a 36\% probability; else, one of the 8 other streams is listened to, each with equal probability.

Every 0.5 to 2 seconds (selected randomly) which represent a \textit{rhythmic loop}, if the value of the selected stream that we call \textit{affect level} is lower than 0.1 the robot arm performs off-page 3-dimensional gestures (\textit{low response}). If the value is between 0.1 and 0.7, one of the 4 gesture types from the language set is randomly selected and executed modulated by the value of the stream (\textit{medium response}). If the value is higher than 0.7, the gesture phrase is interrupted to emulate the startle in the system (\textit{high response}). In addition to that, when the xArm is used with an array of 4 pens, whenever an off-page 3-dimensional gesture is performed the pen currently in use is changed randomly.

Once the phrase is finished or interrupted, this process is repeated to produce the next phrase.

\subsubsection{Execution}
For executing the system, first, the duration of the piece is selected, usually between 4 and 6 mins. Physiological sensors are placed on Jess, the microphone is turned on and the pen (if any) is adjusted. The robot arm first performs an initiating sequence where it will point both sides of its maximum horizontal extent (much like a conductor of an orchestra tapping on their stand to get the orchestra's attention), and then start its process according to the gesture manager. Finally, when the piece is finished, the robot performs a terminating sequence, similar to the initiating one.

The system is started from a main script, enabling the adjustment of different settings including the duration of the piece, sensors and robot used, possible streams used by the gesture manager, and other parameters like enabling automatic termination of the piece when no audio is detected.  \textit{Jess+} is a Python 3 software available in open source on GitHub under the BSD-2-Clause licence \footnote{\url{https://github.com/DigiScore/jess_plus}}.


\section{Results}
\label{workshops}
The three musicians at the core of this experiment were: Jess, a disabled musician with many years of playing experience, who regularly performs electronic music using Ableton Live and a modified wheelchair joystick that sends MIDI data to a laptop; Clare, an experienced professional orchestral violinist, who regularly performs with chamber and small ensembles and works frequently in special education schools on music and creativity workshops; Deirdre, similar experiences to Clare but plays the cello. These three musicians had previously worked together on another music project, but this didn't involve graphic score and improvising. However, they did know each other musically, and this reduced the amount of variables to study.

Four workshops were conducted with the musicians over the 4-month lifespan of the project. A final "sharing" event was scheduled with a closed audience of stakeholders from the partner organisations (6 people in total). The priority during these workshops was to build trust between the team, and to create a sense of trustworthiness with the AI, its behaviour and its growth as an embodied agent in their musicking. There was a balance between improvising, testing, reflecting, and constructive critical developments. Throughout each of these workshops, we ensured that the musicians felt part of the technical development. To achieve this, technical developments and difficulties were expressed using clean language free from jargon and technical complexity. Sometimes this meant agreeing on terms that they could understand and relate to as musicians, such as "temperature" meaning intensity of control parameters. The priority here was to allow their ideas and wishes into the development of their robot regardless of knowledge and experience with computing. Furthermore, the aim was to build the \textit{belief system} and the musical aesthetic of the digital score through their experiences and desires.

\begin{figure}
    \centering
    \includegraphics[width=0.6\linewidth]{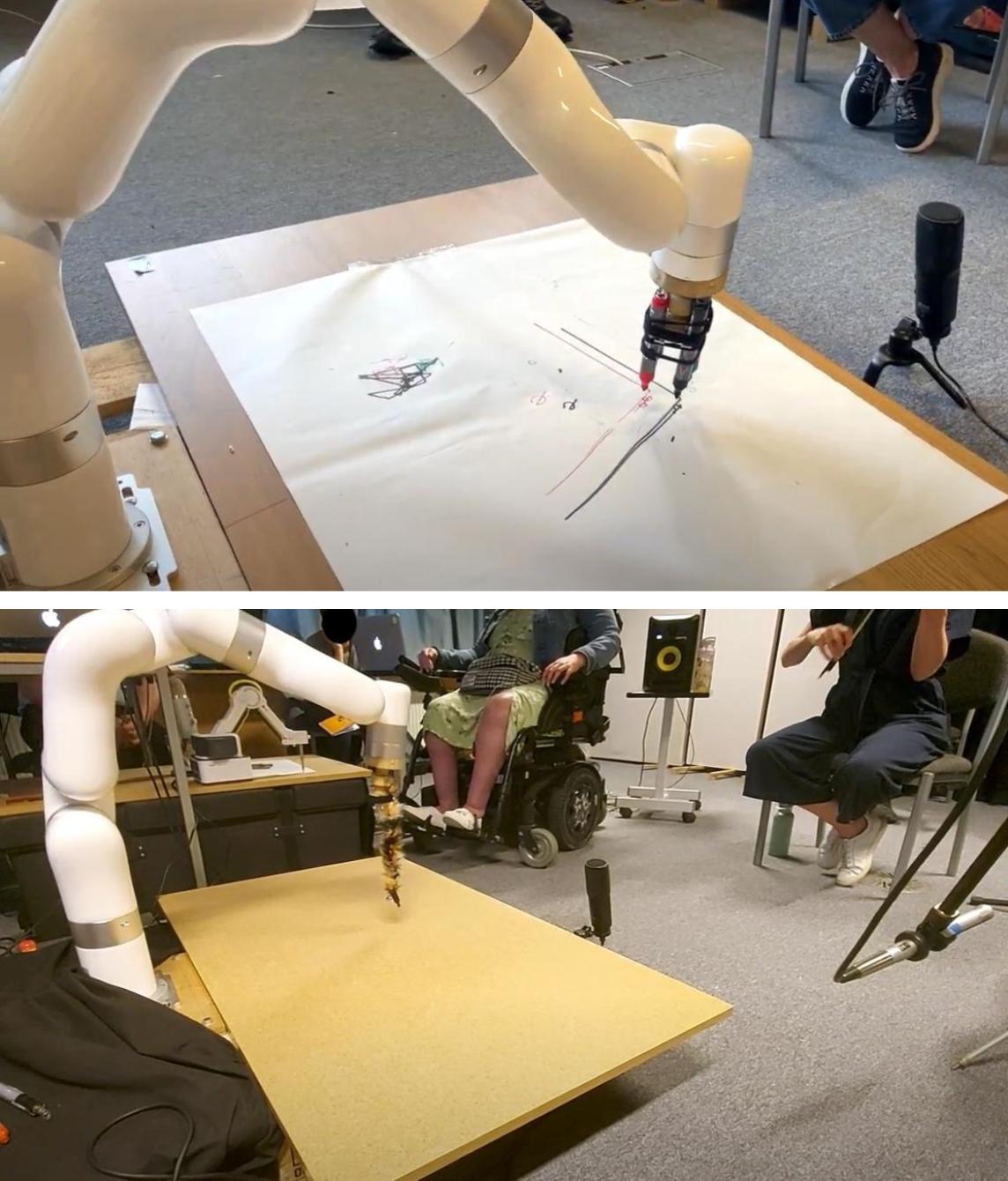}
    \caption{Photos of the final performance}
    \label{fig:final-perf}
\end{figure}

The final session marked the performance in front of an audience for the first time (stakeholders from the partner organisations, 6 people in total). After rehearsals in the morning, the 3 musicians performed in the afternoon four 4-minute-long improvisation pieces. This was the first time that the music had been performed to anyone other than the small development team (the principal investigator, a software developer and a research associate). The musicians felt slight nerves, but were more excited to express to their managers and producers, how the system worked and how it made them feel.
The sharing performance demonstrates the success of the development of the system with this sequence of improvisations:
\begin{itemize} 
    \item[1)] performed with the Magician Lite drawing and no wearable sensors;
    \item[2)] performed using the xArm drawing with 4 pens (top photo in Figure \ref{fig:final-perf}) with wearable sensors \footnote{\url{https://youtu.be/MBPQNmAXvXk}};
    \item[3)] performed using the xArm with a feather (bottom photo in Figure \ref{fig:final-perf}) with wearable sensors \footnote{\url{https://youtu.be/7dQKIpjKJu4}};
    \item[4)] performed using the xArm with a feather (bottom photo in Figure \ref{fig:final-perf}) without wearable sensors \footnote{\url{https://youtu.be/sK4KAmv3ikw}}. 
\end{itemize}

\section{Discussion}
Throughout the extended and iterative research process, we conducted a comprehensive qualitative investigation into the musician's reflections on using this system. This is discussed in detail in a dedicated paper \cite{vear2024jess}, but the analysis of its findings is used here to discuss central aspects of the technical design. As mentioned above the musician's thoughts and reflections were key to developing \textit{Jess+} as the technical developers needed to bring the behaviours of the embodied AI and its interactions inside musicking, into the code of the system. As such, it makes sense to discuss the project from their perspective.

\subsection{Embodied interactions}
In the qualitative research, the musicians discussed their engagement with  \textit{Jess+} as an embodied system. They highlighted how the system operated with them inside music-making, and how it inspired and offered appropriate musical gestures for them to interpret through the flow of the music-making. However, each musician had built a different relationship with the robot. Jess, the disabled musician, discussed how she felt a connection with the system within music-making, and accepted the proposition that \textit{Jess+} was an extension of her, but the purpose of this extension was to draw a creative visual representation of the music being made. This to her was an illustrative artwork of the ensemble's improvisations, using her physiological and psychological data with the collective audio stream that generated the drawing. She referred to it on several occasions, as a "friend", and also as a "story-teller". For the non-disabled musicians, they acknowledged the connection between Jess and the system, and also called it a "friend" but felt that it was more of a "creative accompanist" in which it made a creative contribution to their improvisation through its movement gestures. All the musicians perceived being in-the-loop with the system, and recognising back-and-forth interaction in real-time.

We feel that the design objectives attuned the system in such a way that it supported embodied interactions within music. However, there is so much about "why" the system generated such interactions that we simply don't know, or even understand. The simplicity of the modular design (the AI stack based loosely on Brook's subsumption architecture) made the system lightweight and easy to modify by replacing modules throughout the process, but cannot account for the breadth of musical material generated by the musicians. The AI factory, built on individual neural networks trained using an embodied musicking dataset, seems to be key, as all other design parameters are fuzzy-based behaviours. But a deep dive into correlations within this dataset only brings to the surface extremely loose causality. The fuzzy-based behaviours were attuned to fit with the musical aesthetics of the ensemble, but the extent to which the musicians did not repeat improvised material but invented again and again, cannot be explained by analysing the fuzzy parameters. Nonetheless, those relationships formed by the musicians through their embodied engagement with \textit{Jess+} did transform their musicianship and creativity, \textit{because} of its embodied interactions.

\subsection{Musicianship and creativity}
The musicians highlighted how they felt being in-the-loop with \textit{Jess+} and how the nature of its interactive contribution to musicking transformed their own practices. Jess felt that the system allowed her to express the emotions that she is sometimes not able to express through her current digital setup. For her, being extended through the system meant that she could feel like she was able to express her feelings directly onto a score. “I wanted to explore that part of me and I wanted – you know, I want my emotions that are in here to get expressed outwardly through that” (Jess). For the non-disabled musicians, they felt that they could take risks and, challenge the system as a collaborative partner. This in turn unshackled their improvisational approach and led to emergent novelty in their playing. Overall, they stated that their improvisational skills improved, and felt like they had gained a lot of confidence that could help in future improvisation sessions, especially in their school outreach projects. The musicians acknowledged that \textit{Jess+} established a 'third space' for creativity that flattened any hierarchy of mobility and enhanced the sense of togetherness and inclusion in music-making. They expressed how playing with the system reduced the feelings of "expectations and judgements" (Claire) that playing with other humans could engender.  

In an impact questionnaire following the final performance, Deirdre summed up all their feelings as: 
"The robot arm was liberating to improvise with as it was non-judgemental. At times it united the three musicians' music, and at other times it could also be independent from us (as we knew it would return to respond to what we later did). This in turn influenced the musicians to start or stop, to \textit{gel} together harmonically or feel the freedom to play outside harmonic or rhythmic frameworks. In my opinion, improvising with a human (especially someone new to you) carries psychological elements that could interfere with making music together, so the robot arm provided the opportunity for freedom of musical and emotional expression that would take much more time to establish and develop between humans".


\section{Conclusion}

\textit{Jess+} is an intelligent digital score system for shared creativity with a mixed ensemble of disabled and non-disabled musicians. It was designed using novel approaches to embodied AI and human-robot interaction. The development of the system was iterative and human-centred, placing 3 core musicians at the centre of the development process. Our findings showed that the design decisions that were implemented through the embodied-AI approach led to rich experiences for the musicians which led to transformations of their practice and creative engagement as an inclusive ensemble.  However, many unknowns have emerged out of this proof-of-concept, that we cannot say with any certainty \textit{how} they led to such transformations encounters for the musicians. This is the next stage of the research, but we feel confident to assert that embedding music-specific embodied interaction data and behaviours into each aspect of the design did contribute to their transformational encounters. We simply need to investigate why.


\section{Author contributions}
CV initially conceptualised the system; CV and JB both contributed equally to the development of the system and the writing of the manuscript.


\section{Acknowledgments}
This work was funded by the ERC under the European Union's Horizon 2020 research and innovation programme (ERC-2020-COG – 101002086). Additional funding was received from TAS Hub and the Faculty of Arts at the University of Nottingham. We would like to address our sincere thanks for the support we received from Orchestras Live\footnote{\url{https://www.orchestraslive.org.uk/}} and Sinfonia Viva\footnote{\url{https://www.sinfoniaviva.co.uk/}}, and of course our musicians Jess, Claire and Deirdre.


\bibliography{aaai}

\end{document}